\documentclass[12pt]{article}
\pdfoutput=1
\usepackage{amsbsy}
\usepackage{amsfonts}
\usepackage{amsmath,amssymb}
\usepackage{bbold}
\usepackage{graphicx}

\usepackage{mathtools}

\newcommand{\sect}[1]{\setcounter{equation}{0}\section{#1}}

\newcommand{\eq}{\begin{equation}}
\newcommand{\eqa}{\begin{eqnarray}}  
\newcommand{\en}{\end{equation}}
\newcommand{\ena}{\end{eqnarray}}
\newcommand{\enn}{\nonumber \end{equation}}

\def\sk{\vskip .4cm}
\def\noi{\noindent}

\def\al{\alpha}
\def\be{\beta}

\let \part\partial

\def\part{\partial}

\def\sk{\vskip .4cm}

\def\noi{\noindent}

\def\X0{X^0}

\def\al{\alpha}

\def\square{{\,\lower0.9pt\vbox{\hrule \hbox{\vrule height 0.2 cm
\hskip 0.2 cm \vrule height 0.2 cm}\hrule}\,}}

\def\lb{\langle}
\def\rb{\rangle}

\begin{document}

\begin{titlepage}

\vskip 2em
\begin{center}
{\Large \bf All quantum mixtures are proper} \\[3em]

\vskip 0.5cm

{\bf
Leonardo Castellani}
\medskip

\vskip 0.5cm

{\sl Dipartimento di Scienze e Innovazione Tecnologica
\\Universit\`a del Piemonte Orientale, viale T. Michel 11, 15121 Alessandria, Italy\\ [.5em] INFN, Sezione di 
Torino, via P. Giuria 1, 10125 Torino, Italy\\ [.5em]
Arnold-Regge Center, via P. Giuria 1, 10125 Torino, Italy
}\\ [4em]
\end{center}

\begin{abstract}
\sk

It is argued that proper and improper quantum mixed states have no observable differences, and hence
should not be distinguished. This has implications for subjective approaches to quantum mechanics, and invalidates one of the main motivations for relational interpretations of QM.

\end{abstract}

\vskip 10cm\
 \noi \hrule \vskip .2cm \noi {\small
leonardo.castellani@uniupo.it}

\end{titlepage}

\newpage
\setcounter{page}{1}


\sect{Introduction}

Measurements by definition presuppose a separation between the observed system S and the observer O. This separation can be moved towards S or towards O, thus redefining S and O, as was realized from the early days of quantum mechanics (the ``Heisenberg cut"). Once the separation is decided, so that subsystems S and O are precisely identified, the quantum mechanical rules 
provide statistical predictions on the results of measurements performed by O on the system S.

Take for simplicity S to be a qubit, i.e. a two-dimensional quantum system, and A to be an observable of S with orthonormal eigenvectors $|0\rb$ and $|1\rb$. If S is in the pure state $|\psi\rb = \alpha |0\rb + \beta |1\rb$,
and O measures A without registering the result, the pure state collapses into a mixture described by the
density matrix 
\eq
|\psi\rb \lb \psi|  \longrightarrow |\alpha|^2 |0\rb \lb 0| + | \beta|^2 |1\rb \lb 1| \label{collapseS}
\en
This statistical mixture indicates that S is in the
state $|0\rb$ with probability $|\alpha|^2$ and in the state $|1\rb$ with probability $|\beta|^2$, and reflects the ignorance of the measurement result. The collapse (\ref{collapseS}) 
is the ``ontic part" of the process, being independent of the knowledge
of the measurement result. The ``epistemic part"  is in the sharper description $|0\rb \lb 0|$ or
$|1\rb \lb 1|$, depending on this knowledge. Thus we can describe the measurement as a two step
process, an ``ontic" collapse followed by an ``epistemic" collapse:
\eq
(\alpha |0\rb + \beta |1\rb)(\lb 0|\alpha^* + \lb 1| \beta^*) \xrightarrow{``ontic" collapse}   |\alpha|^2 |0\rb \lb 0| + | \beta|^2 |1\rb \lb 1|  \label{ontic}
\en
\eq
 |\alpha|^2 |0\rb \lb 0| + | \beta|^2 |1\rb \lb 1| \xrightarrow[(knows~ that ~result ~is ~1)]{``epistemic" ~collapse} |1\rb\lb1| \label{epistemic}
\en

In what follows, our concern will be with the ontic part, since it is the genuinely quantum mechanical part
of the measurement process. The epistemic part is essentially the same as in classical situations, for example
when flipping a coin and looking or not looking at the result.

Here we assume that all physical systems, including observers, are described by quantum mechanics, and that ``classicality" arises as an emergent phenomenon. By ``observer" we mean a physical system (human or not) that can interact with S, and due to this interaction becomes correlated with S. By ``measurement" we mean
this interaction, occurring at a definite time, and producing the ontic collapse.

We also assume that the state of a system is {\sl completely} described by its density matrix, as introduced in \cite{vonNeumann1932} (see also \cite{Landau1927}). In particular we consider the state of a subsystem in a composite system to be  {\sl completely} described by its reduced density matrix.

\sect{Measurement as interaction}

A central tenet of QM is the unitary evolution of an isolated system S, according to the Schr\"odinger equation.  On the other hand, the measurement of an observable on S implies an interaction with an external measuring apparatus O: the system S in this case is {\sl not} isolated, and its evolution during the measurement is described by a (nonunitary) projection. The clash between unitary evolution ({\sl Process 2} in von Neumann terminology \cite{vonNeumann1932}) and projection ({\sl Process 1}) is at the core of the so called ``measurement problem" in quantum mechanics (see for ex. \cite{Schlosshauer,Brukner} and ref.s therein). We will argue that the problem disappears if we consider only the ontic part of the collapse, as described by (\ref{ontic}) in the previous Section, where a pure state collapses into a mixture. The epistemic part we consider irrelevant to the discussion, since it is intrinsic to all statistical ensembles (not necessarily quantum). 

More precisely, the problem seems to arise when we consider the (isolated) composite system S+O and want to describe its evolution during the measurement. The composite system, being isolated from external influences, must evolve unitarily. But how can we describe unitarily a measurement that takes place inside S+O, involving a nonunitary process on the subsystem S ? The solution is provided by the reduced density matrix formalism. Indeed the reduced density matrix 
for the subsystem S, after the interaction with the measuring apparatus O,  describes {\sl exactly the same statistical mixture} as in (\ref{collapseS}), with weights equal to the Born probabilities. This is a remarkable feat of the formalism of textbook QM, and allows the interpretation of S as a ``measured system" within a purely unitary evolution.

To be concrete, consider S and O to be two qubits, i.e. two-dimensional quantum systems, each with orthonormal basis vectors $|0\rb$ and $|1\rb$. Moreover define observables A on S and B on O, both with eigenvectors given by $|0\rb$ and $|1\rb$. The unitary evolution must entangle S with O, since states of S must be correlated with states of the apparatus O. This can be realized with a CNOT gate, i.e. a unitary operation on the composite system acting on the basis vectors as
 $|0\rb|0\rb \rightarrow
|0\rb|0\rb$, $|0\rb|1\rb \rightarrow
|0\rb|1\rb$, $|1\rb|0\rb \rightarrow
|1\rb|1\rb$, $|1\rb|1\rb \rightarrow
|1\rb|0\rb$.

Consider as initial state of S+O the product state $(\alpha |0\rb + \beta |1\rb) |0\rb$. Then the CNOT gate transforms it, by linearity, into the entangled state $\alpha |0\rb |0\rb  + \beta |1\rb |1\rb$, cf. Fig. 1. Thus, results of measurements of A on S are correlated with those of B on O. 
  \sk
\includegraphics[scale=0.45]{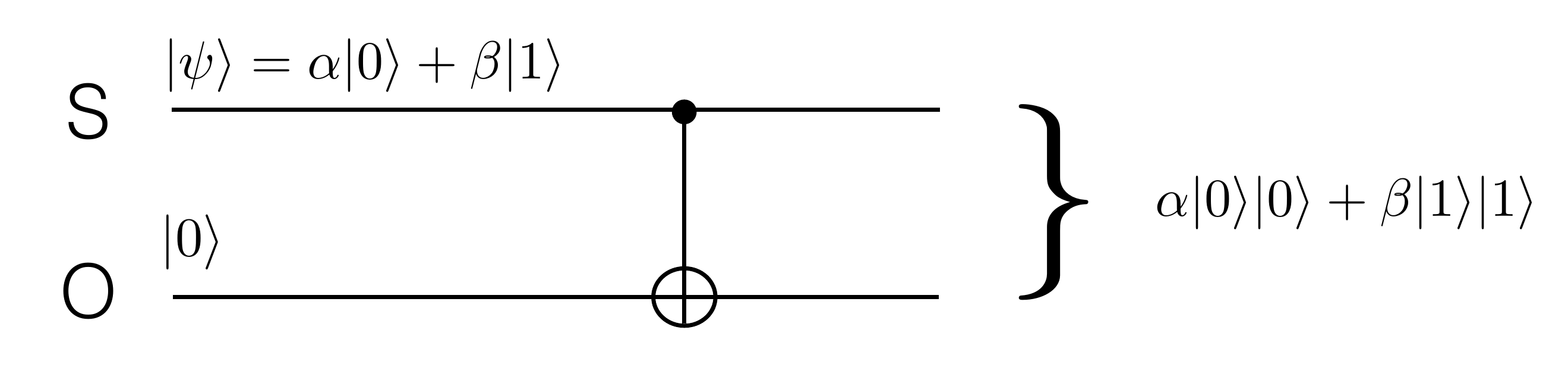}

\noi {\bf Fig. 1}  {\small Circuit modeling the measurement of an observable with eigenvectors $|0\rb$ and $|1\rb$. Both the measured system S and the observer (or apparatus) O are taken to be 1-qubit systems. }  
\sk

Is this a good model for quantum measurements, if B is identified with the pointer position of a measuring apparatus O ? At first sight there seems to be a problem,
since the effect of a measurement on S should project it into one of the two base states $|0\rb, |1\rb$, with probabilities $|\alpha|^2$ and $|\beta|^2$ respectively. Here instead we have a final entangled state for the composite system. How can we resolve this apparent disagreement ?

The answer is to focus on the state of the {\it subsystem} S. This state is completely described by the reduced density matrix $\rho^S$, obtained by tracing the density matrix of S+O over system O. The partial trace yields
\eq
\rho^S = Tr_O [ (\alpha |0\rb |0\rb  + \beta |1\rb |1\rb)(\alpha^* \lb 0 | \lb 0|  + \beta^*  \lb 1|  \lb 1|)] =
|\alpha|^2 |0\rb \lb 0| + | \beta|^2 |1\rb \lb 1|  \label{mixedstate}
\en
and describes the state of subsystem S after the CNOT interaction. Note that $\rho^S$ describes a {\it mixed state} of S. Before the CNOT interaction, S was in the {\it pure state} $\alpha |0\rb + \beta |1\rb$.  Therefore the effect of the interaction on the subsystem S has been
a {\sl nonunitary transformation}, even if S+O evolves unitarily.

 The statistical mixture (\ref{mixedstate}) coincides with (\ref{collapseS}).  Thus the ontic part of the measurement process can be described by a unitary evolution of S+O. 
 \sk
 \noi {\bf Note:} the quantum circuit in Fig. 1 implements a measurement in the computational basis also on {\sl mixed} states. Indeed we can replace the pure state $|\psi\rb$ (input state for the first qubit) with
 an input density matrix $\rho_1$, while keeping the state of
 the second qubit equal to $|0\rb$. Take for example $\rho_1 = p_a |a\rb \lb a| + p_b |b\rb \lb b|$  with $p_a+p_b=1$, the generalization to an arbitrary ensemble being straightforward. The pure states $|a\rb,|b\rb$ do not need to be orthogonal, and are expanded on the computational basis as
 \eq
 |a\rb = \alpha |0\rb + \beta |1\rb, ~~~|b\rb = \gamma|0\rb + \delta |1\rb
 \en
 Then the density
 matrix $\rho$ for the 2-qubit system evolves as
 \eq
 \rho = (p_a |a\rb \lb a| + p_b |b\rb \lb b|) \otimes |0\rb \lb 0| \longrightarrow \rho' = {\rm CNOT}~\rho~{\rm CNOT}  \label{rhoprime}
 \en
 under the action of the circuit in Fig. 1. Recalling that CNOT = $|0\rb \lb 0| \otimes I + |1\rb \lb 1| \otimes X$ where $X$ is the quantum NOT gate, i.e. $X|0\rb = |1\rb, X|1\rb = |0\rb$, it is an easy exercise to find
$ \rho' $ as given by (\ref{rhoprime}). Taking then its partial trace  
yields the reduced density matrix $\rho'_1$ for the first qubit:
\eq
\rho'_1 = Tr_2 (\rho') = (p_a |\alpha|^2 + p_b |\gamma|^2) |0\rb\lb 0|  + (p_a |\beta|^2 + p_b |\delta|^2) |1\rb\lb 1|
\en
 This is exactly the same density matrix one obtains for the first qubit after measuring it (and ignoring the result),  using the textbook rule
 \eq
 \rho'_1 = \sum_m P_m \rho_1 P_m
 \en
$P_m$ being the projectors $|0\rb\lb 0|$ and $|1\rb\lb 1|$. The transformation $\rho_1 \rightarrow \rho'_1$ is in general not unitary (quick proof: $Tr (\rho'_1)^2$ in general differs from $Tr (\rho_1)^2$), and represents the ontic collapse of the density matrix under a measurement. Thus the ontic part of the measurement  is described by a unitary evolution of the S+O density matrix.

 \sect{Objections}
 
 An often heard objection \cite{dEspagnat1976,Hughes1989} relies on the distinction between proper and improper mixtures, a terminology introduced by D' Espagnat \cite{dEspagnat1976}.  A statistical mixture
 due to ignorance is called proper, whereas the same mixture describing a subsystem of a composite system in an entangled pure state is called improper. In the example of the preceding Section, the objection runs as follows. We cannot interpret the (improper) statistical mixture (\ref{mixedstate}) to describe a system which is in one of the two states $|0\rb, |1\rb$, with probabilities $|\alpha|^2$ and $|\beta|^2$ respectively: if this were the case, the total system should find itself in one of the two states $|0\rb |0\rb$, $|1\rb |1\rb$ with probabilities $|\alpha|^2$ and $|\beta|^2$ respectively, implying for its density matrix to be
 \eq
|\alpha|^2 |0\rb |0\rb \lb 0| \lb 0|  + |\beta|^2 |1\rb |1\rb \lb 1| \lb 1|  \label{density1}
 \en
 in contradiction with the density matrix of S+O after the CNOT interaction
 \eq
 (\alpha |0\rb |0\rb  + \beta |1\rb |1\rb)(\alpha^* \lb 0 | \lb 0|  + \beta^*  \lb 1|  \lb 1|) \label{density2}
 \en
 
 Another objection invokes ``basis ambiguity", meaning that the same density matrix $\rho$  can correspond to different statistical ensembles (for an exhaustive study see \cite{Hughston1993}). Therefore knowing $\rho$ does not imply knowing which observable has been measured. For example in the case $\alpha=\beta=1/\sqrt{2}$, the reduced density matrix for S in (\ref{mixedstate}) becomes 
 \eq
 {1 \over 2} ( |0\rb \lb 0| + |1\rb \lb 1| ) = {1 \over 2} ( |a\rb \lb a | + |b\rb \lb b| ) \label{basisambiguity}
 \en
 for any orthonormal vectors $|a\rb, |b\rb$ connected to $|0\rb,|1\rb$ by an orthogonal transformation.
 Thus the same reduced density matrix would describe the state of S after a measurement, where the measured observable could be one of an infinite set of mutually incompatible (noncommuting) observables.  
 
 Note that the same basis ambiguity holds for the state of the composite S+O system after the CNOT interaction, when $\alpha=\beta=1/\sqrt{2}$. This is often referred to as the ``preferred-basis problem".

 \sect{Counterobjections}
 
 Here we counter both objections, on logical and physical grounds. There have been numerous rebuttals and counter-rebuttals over the years, see for ex. \cite{Anandan1998,dEspagnat1998,Kirkpatrick2001,dEspagnat2001}, and most of the present Section is far from original. If the 
 density matrix $\rho^S$ gives a {\sl complete} description of the state of S, it makes no sense to distinguish
 states corresponding to the same $\rho^S$. If only S is accessible to observations, no experiment can
 distinguish between a proper or an improper mixture. The fact that S is part
 of a larger system S+O, and that this larger system is in a pure entangled state, can only be revealed
 by measurements on the {\sl whole} S+O. Only then one can discriminate between the two states of the composite
 system S+O, given in (\ref{density1}) and in (\ref{density2}), respectively a pure entangled state and a statistical mixture. These two different states give rise to the
 same reduced density matrix for S, and this matrix defines a {\sl unique} physical state, i.e. a statistical mixture,
 neither proper nor improper. All measurements made on S are consistent with the ignorance interpretation,
 and therefore $\rho^S$ in (\ref{mixedstate}) has all the markings of a state of a measured system.
 \sk
The second objection in the preceding Section questions the interpretation of the mixed state (\ref{basisambiguity}) as the state of a measured system, since it seems that in this case the measured observable cannot be specified uniquely,\footnote{more precisely, its eigenvectors cannot be specified uniquely.} due to basis ambiguity. To this we answer as follows:
 
 i) Basis ambiguity does not compromise in any way the interpretation of (\ref{basisambiguity}) as the state of a measured system: it may very well be the state obtained after measuring any of a whole family of possibly noncommuting observables. The fact that the measurements of any of those observables lead to the same statistical mixture does not mean that the mixture does not describe correctly the state after the measurement.
 
 ii) We can assume that the observer knows which observable is being measured, i.e. what particular
 interaction takes place between the apparatus and the system. In our simplified two-qubit example this interaction is given by a CNOT gate, and models a measurement of an observable in the computational basis, i.e. of an observable with eigenvectors $|0\rb$ and $|1\rb$. This removes the basis ambiguity: the basis, in which the mixed state (\ref{mixedstate}) is to be interpreted as a state of a measured system, is the computational one. It is really the measuring apparatus, with its specific interaction with the system, that dictates the ``interpretation basis". If the observer wants to measure a different observable, for example with eigenvectors $|+\rb$ and $|-\rb$, then the circuit implementing the system-apparatus interaction must be modified as follows:
 \sk
 \sk
 
\includegraphics[scale=0.40]{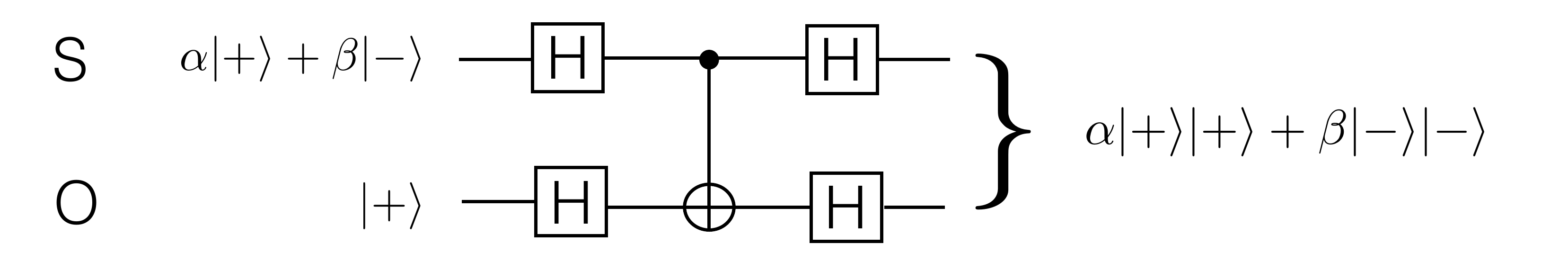}

\noi {\bf Fig. 2}  {\small Circuit modeling the measurement of an observable with eigenvectors $|+\rb$, $|-\rb$. The gate $H$ is the Hadamard gate, defined by $H |0\rb = |+\rb$,  $H |1\rb = |-\rb$ .}  
\sk

\noi In this case, with $\alpha=\beta=1/\sqrt{2}$, the reduced density matrix for the first qubit, after the interaction, is $\rho^S = 1/2 (|+\rb \lb +| + |-\rb \lb -|)$, equal to the one obtained in the circuit of Fig.1, i.e.  $\rho^S = 1/2 (|0\rb \lb 0| + |1\rb \lb 1|) $. However the interpretation basis is different, and is given by the orthonormal couple $|+\rb,|-\rb$. Indeed the new circuit in Fig. 2 correlates the states $|+\rb$,  $|-\rb$ of the system to the states $|+\rb$, $|-\rb$  of the apparatus, respectively, and thus models the measurement of an observable with $|+\rb$ and $|-\rb$ eigenvectors. Thus the choice of an interpretation basis is not ambiguous, but can be considered epistemic, since it relies on the observer's knowledge of which observable is being measured.

 \sect{Refinements}
  
  With the same logic, we can describe a measurement on a spatially extended
  2-qubit system S in the entangled state ${1 \over \sqrt{2}} (|0\rb |0\rb  + |1\rb |1\rb)$, where now 
  the observer (call her Alice) has access only to the first qubit. Standard quantum mechanics tells us that
  Alice has a probability $1/2$ to obtain 0 or 1, and that the measurement produces a corresponding collapse
  of S into one of the states $|0\rb |0\rb$  or $|1\rb |1\rb$.  Again we can describe the ontic part of this measuring     
  process by a specific interaction between Alice and her qubit, the only accessible (to Alice) part of S. Modeling  
  Alice as a single qubit, the interaction is given by the same CNOT gate used in Section 2:

  \sk
\includegraphics[scale=0.45]{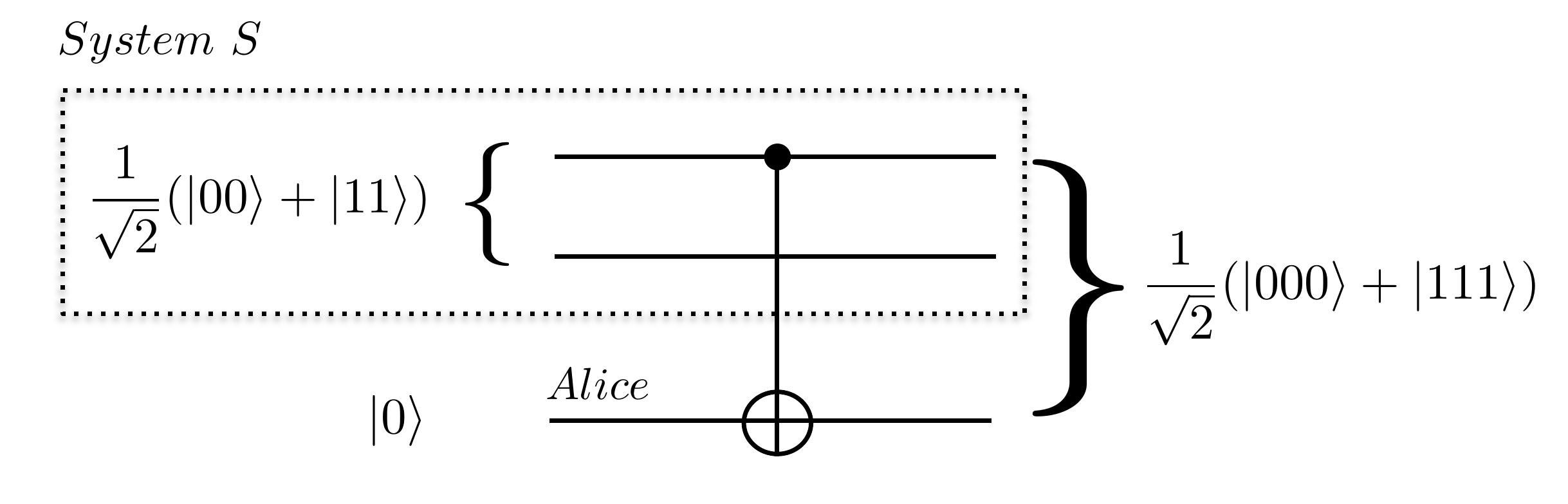}

\noi {\bf Fig. 3}  {\small Circuit modeling the measurement by Alice on an entangled state of system S, where she has access only to the first qubit. She interacts with the first qubit via a CNOT gate. }  
\sk
\noi The state of Alice is initialized to $|0\rb$ (the ``ready state"). The output is the maximally entangled 3-qubit state given in Fig. 3.  To find the state of the subsystem S after the interaction one must trace over A the total density matrix, which yields the reduced density matrix for S:
  \eq
  \rho^S = Tr_A (\rho^{SA}) = {1\over 2}Tr_A (|000\rb+|111\rb)(\lb 000| + \lb 111|)={1\over 2} |00\rb \lb 00| + {1\over 2} |11\rb \lb 11| 
  \en
  This reduced matrix describes system S as a system that has collapsed to the state $|00\rb$ or $|11\rb$ with probability $1/2$: this is the ``ontic" collapse, and occurs precisely at the time of the interaction. If Alice ``looks" at the result after the measurement, she will know whether the system S is in the state $|00\rb$ or in the state $|11\rb$, thus producing the ``epistemic" collapse. This part of the collapse occurs instantaneously on the whole extended system S, without any paradox since it is related to a ``state of knowledge" of Alice.  
  
  We can adapt this discussion to a Schr\"odinger's cat situation, imagining that the
  entangled initial state of S has been produced by an interaction between a decaying atom and the cat.
  Then Alice ``measures the cat", in the sense that her state becomes entangled with the cat state ($|alive\rb$ or $|dead\rb$). This ontic measurement produces the final state in Fig. 3. Tracing over Alice's degrees of freedom yields a mixed state for the subsystem cat + atom. This mixed state is a measured state, not
  a superposition state: the cat is either alive or dead, with probabilities depending on the particular
  interaction between the cat and the atom (in Fig. 3 these probabilities are equal).  Thus no
  superposition of $|alive\rb$ and $|dead\rb$ states is left, and the ``paradox" disappears.

     \sect{A critique of state subjectivism}  
  
  In most subjectivist or relational interpretations of QM  (see for ex. \cite{Qbism1,Qbism2,Everett,Rovelli}) the alleged {\sl observer dependence} of quantum states is claimed to be a consequence of the usual quantum mechanical rules, when these are applied to ``third person" situations, as in Wigner's friend thought experiment \cite{Wigner1961}. This involves an isolated lab L, where an observer F (the ``friend")  measures a system S, and an external observer W  (``Wigner") outside of L. It is claimed that F and W give different descriptions of the state of  S after a measurement by F, essentially because for F the system S has collapsed into a projected state (eigenvector of the measured observable) whereas for W the whole system in L, being isolated,  must evolve unitarily.  Again the contradiction is only apparent, since W and F should compare their
  descriptions of the {\sl same} system S: if they do so, they will both agree that S, after the measurement,
  is in a mixed state, for the same reason discussed in Section 2. Modeling S and F as one-qubit systems, and (the ontic part of) the measurement as a CNOT gate, the reduced density matrix for S describes its state as the same mixture for {\sl any} observer. If, in addition,  F also knows the result of the measurement, he will be able
  to describe the state of S as a particular state of the mixture. This further sharpening, due to additional knowledge of F (W has no access to the inner contents of L), is what produces the ``epistemic collapse",
  common to all situations (classical or quantum) where one has incomplete information on the state of a system, and encodes this ignorance in a statistical ensemble.
  
   To defend their thesis, supporters of the non-objectivity of quantum states often insist on the distinction between proper and improper mixtures, and on the claim that only proper mixtures can describe the state of a measured system (with ignorance of the result).  One can understand why this is a very sensitive point in any
  discussion regarding the ``objectivity of the wave function". In Section 4 we have argued that the proper/improper distinction is unphysical, i.e. not detectable by any experiment on S. Therefore we conclude that Wigner's friend experiment cannot be taken as cornerstone of a relational interpretation of QM, as done for example in \cite{Rovelli}. 
  
  \sect{Conclusions}
 
  We have presented a case for the objectivity of quantum states, further elaborating and developing some arguments discussed in \cite{LC}. For this we have recalled the description of a measured system as a mixture,  coinciding with the one obtained by tracing over the observer's degrees of freedom. The only epistemic ingredient of a quantum mixture is due to the incomplete information, encoded in the statistical ensemble of the mixture. This ingredient
  is intrinsic to the very definition of a statistical ensemble, and is not a characteristic of the quantum world.
  For the whole argument we used the standard rules of QM: in particular the reduced density matrix formalism
allows to describe measurement as an unitary process, i.e an interaction
  between observer and system. In our opinion many of the so-called paradoxes 
  or interpretative problems of QM are 
  resolved by making use of the full power of its formalism. In other words, QM is quite
  capable to cure itself from apparent internal contradictions.

 \section*{Acknowledgements}

This work is supported by the research funds of the Eastern Piedmont University and
INFN - Torino Section.

\vfill\eject

\begin{thebibliography}{99}

\bibitem{vonNeumann1932}
J. von Neumann, ``Mathematical Foundations of Quantum Mechanics,“  Princeton University Press, Princeton, N. J., 1955, R.T. Beyer, tr; originally published as Mathematische Grundlagen der Quantenmechanik, Springer, Berlin, 1932.

\bibitem{Landau1927} 
L. Landau, ``Das D\"ampfungsproblem in der Wellenmechanik", Z. Phys. {\bf 45} (1927) 430-441.

\bibitem{Schlosshauer}
M.~Schlosshauer,
``Decoherence, the Measurement Problem, and Interpretations of Quantum Mechanics,''
Rev. Mod. Phys. \textbf{76}, 1267-1305 (2004)
[arXiv:quant-ph/0312059 [quant-ph]].

 \bibitem{Brukner}
C. Brukner, ``On the quantum measurement problem." In R. Bertlmann and A. Zeilinger, editors,
Quantum [Un]Speakables II: Half a Century of Bell’s Theorem, pages 95–117. Springer, 2017;
arXiv:1507.05255.

\bibitem{dEspagnat1976}
B. d’Espagnat, ``Conceptual Foundations of Quantum Mechanics," 2nd ed., W. A. Benjamin, Menlo Park, CA, 1976.

\bibitem{Hughes1989}
R. I. G. Hughes, ``The Structure and Interpretation of Quantum Mechanics," Harvard University Press, Cambridge, MA., 1989.

\bibitem{Hughston1993}
L. Hughston, R. Jozsa and W. Wootters,  ``A Complete Classification of Quantum Ensembles Having a Given Density Matrix.“ Physics Letters A, 183, 14 - 18 (1993).


\bibitem{Anandan1998}
J.~Anandan and Y.~Aharonov,
``Meaning of the density matrix,''
Found. Phys. Lett. \textbf{12}, no.6, 571-578 (1999)
doi:10.1023/A:1021699226154
[arXiv:quant-ph/9803018 [quant-ph]].

\bibitem{dEspagnat1998}
B. d'Espagnat,  Reply to Aharonov and Anandan's  ``Meaning of the Density Matrix",  arXiv:quant-ph/9804063, 1998.

\bibitem{Kirkpatrick2001}
K. A. Kirkpatrick,  ``Indistinguishability and improper mixtures", arXiv:quant-ph/0109146 (2001)

\bibitem{dEspagnat2001}
B. d' Espagnat, ``Reply to K A Kirkpatrick," arXiv:quant-ph/0111081, 2001

\bibitem{Qbism1} C. A. Fuchs, R. Schack, ``Quantum-Bayesian coherence,"
Rev. Mod. Phys. 85, 1693 (2013).

\bibitem{Qbism2}  N. D. Mermin, ``Physics: QBism puts the scientist back
into science," Nature 507, 421–423 (2014).

\bibitem{Everett} H. Everett, ``Relative state formulation of quantum mechanics,"  Rev. Mod. Phys. 29, 454–462 (1957).

\bibitem{Rovelli} C. Rovelli, ``Relational Quantum Mechanics," International Journal of Theoretical Physics {\bf 35} (1996) 1637; arXiv:quant-ph/9609002.

\bibitem{Wigner1961} E. P. Wigner, in The Scientist Speculates (ed. Good, I. J.) 284–302
(Heinemann, 1961).

\bibitem{LC}
L.~Castellani,
``No Relation for Wigner\textquoteright{}s Friend,''
Int. J. Theor. Phys. \textbf{60}, no.6, 2084-2089 (2021).


 \end{thebibliography}
\end{document}